## Contents

'Uchitel' Publishing House  Moscow

# The Rohingyas of Rakhine State: Social Evolution and History in the Light of Ethnic Nationalism

**Sarwar J. Minar**
*Northern Illinois University*

**Abdul Halim**
*Journalist and Independent Researcher*

**ABSTRACT**

*In August 2017, over 725,000 Rohingya Muslims and Hindus were ousted from Rakhine State by the Tatmadaw as it undertook a brutal attack in response to Arakan Rohingya Salvation Army's coordinated attacks on a military base and security force outposts resulting death of twelve. The UN finds Tatmadaw had sheer intent of 'ethnic cleansing' and referred the attack as a 'textbook example' of ethnic cleansing. These events provoked worldwide public-and-academic interest in history and social evolution of the Rohingyas, and this is to what the article is devoted. As the existing literature presents a debate over 'Who are the Rohingyas?', and 'How legitimate is their claim over Rakhine State?', the paper reinvestigates the issues using a qualitative research method. Compiling a detailed history, the paper finds that Rohingya community developed through historically complicated processes marked by invasions and counter-invasions. The paper argues many people entered Bengal from Arakan before British brought people into Rakhine state. The Rohingyas believe Rakhine State is their ancestral homeland and they developed a sense of 'Ethnic Nationalism'. Their right over Rakhine State is as significant as other groups. The paper concludes that the UN must pursue solution to the crisis and the government should accept the Rohingyas as it did the land or territory.*

## 1. INTRODUCTION

Who are the Rohingyas? How legitimate is their claim of belongingness in the Rakhine State in Myanmar? The conventional wisdom presents a moderate debate over claims and counter claims: one group of scholars (*e.g.*, Derek Tonkin, Jacques Leider *etc*.) claims Rohingyas are the people who entered Myanmar during the British rule, and another group (*e.g.*, A. P. Phayre, G. E. Harvey, M A Tahir Ba Tha, Abdul Karim, Azeem Ibrahim, Kazi Fehmida Farzana *etc*.) claims that Rohingyas have been living in Rakhine State for centuries. In the backdrop of the debate, this article reinvestigates the issues through qualitative research method assessing the perspectives of the ousted Rohingyas from Myanmar who took shelter in Bangladesh and reevaluating the secondary literatures and historical documents.

Rakhine State has historically been known as Arakan. Though the history of the Rohingyas' inflow into Arakan often referred to the time of British annexation of Burma, this article delves deeper and finds that the history prior to British annexation of Arakan and Burma has been very significant. Though there is no denying the fact that during British colonial period many people were brought into Rakhine state from then Bengal (mostly from Chattogram) and India, it is less known that a huge number of people fled from Arakan/Rakhine State to Chattogram (Bengal) during Burmese attack on Arakan prior to British annexation of Arakan/Burma forming a significant part of the base population in Chattogram, Bangladesh. Similarly, people are less aware that the history of Arakan is marked by many such instances of flow of people. It is imperative that we take the history of such flow of people into serious consideration.

The article is organized into eleven sections. Following the introduction, the second section outlines the research methodology and theoretical framework. Third section briefly summarizes the literature review. The fourth section presents the argument of Rohingyas as a case of ethnic nationalism. The historical evidences are presented from fifth to ninth sections. The fifth section conceptualizes Arakan as a region and draws the history of Arakan in ancient time. The sixth describes the early infiltration of various groups into Arakan, and the seventh section portrays history of various invasions and counter-invasions between and among Arakan and its neighboring political entities. The eighth section brings up the developments in Arakan when it was under British Crown Colony, and the ninth section draws insights in the post-independent period of Myanmar. The tenth section draws the social evolution of the Rohingyas in Rakhine State during pre-colonial, colo-nial, and postcolonial period. Lastly, the eleventh section discusses the findings and draws conclusion.

## 2. RESEARCH METHODOLOGY AND THEORETICAL FRAMEWORK

This is a qualitative research project and uses nonparticipant observation method relying on both primary and secondary data. The primary data are collected from fifty interviews in the field using a semistructured questionnaire on the Rohingyas living in Teknaf, Cox's Bazar, Bangladesh. The secondary data are collected from various available secondary literature that includes books, journal articles, reports, newspapers, blogs, news portals and so forth. Based on the primary data from the field interviews, the paper assesses how the Rohingyas see themselves or their identity and how they relate their claim over their living place in Rakhine State, Myanmar. In order to understand such historical context, the article uses the secondary literature on history to study Rohingyas' history, because the secondary literatures serve as the best data source when there is no other better available data and this definitely enriches the research.

Based on the findings of the interviews, the article appropriates evolutionary approach to ethnic nationalism and ethnic nationalist identity formation of the Rohingyas to address the issue. 'Ethnic nationalism' is a form of nationalism where nation is defined in terms of the ethnicity of the people, whereas, 'nation' may be defined as unity of people based on language, shared culture, history, heritage, a common and shared ethnic ancestry *etc*. (Yun 1990).

The article uses 'Ethnic Nationalism' as theoretical framework which was developed by Anthony D. Smith (Smith 1994) as it is considered as the most satisfactory among the three spectra of nationalism: Ethno Nationalism, Ethnic Nationalism and Mini Nationalism (Yun 1990). Smith provides a comparison between 'Territorial Nationalism' and 'Ethnic Nationalism.' While 'Territorial Nationalism' denotes to formation of nationalism based on a territory, the living place of its inhabitants, which gives a sense of bondage and common identity that goes beyond other lesser bondages, and assimilates ethnic, linguistic, religious, cultural differences, 'Ethnic Nationalism' instead focuses on the ethnicity of the people which gives a sense of bondage and common identity. However, the sense of belongingness may not always be something inherent rather it may also be created by the elite class and imposed on the general people for greater political leverage.

## 3. LITERATURE REVIEW

Numerous research have been undertaken in this area. Jacques P. Leider in *Rohingya, Rakhine and the Recent Outbreak of Violence-A Note* (2012) and *Rohingya: the Name, the Movement and the Quest for Identity* (2014) argued that there is no doubt of historical existence of Muslim community in Rakhine; however, in true sense they are not the descendants of pre-colonial Muslim community, rather the Rohingyas are the Bengali originated Muslim population of Rakhine State. He also argued that the claim of early settlement on Rohingyas ancestor in the seventh century by Arab sailorman is a matter of Sheer believes. A British diplomat, Derek Tonkin, in *The 'Rohingya' Identity: British Experience in Arakan 1826–1948* (2014) argued that the dialect Rohingya is a recent construction of Bengali speaking Muslim of Rakhine. He claimed that the word 'Rohingya' had never been used as an ethnic identity rather it was a political tag to define Muslim majority in Rakhine after World War II.

Sir Arthur Purves Phayre in *History of Burma* (1883), extensively illustrated the pre-historic and early historical developments of Arakan. He pointed out the pre-historic settlements and initial developments in Arakan till the British anexation. Godfrey Eric Harvey in *History of Burma* (1925) also depicted the history of Burma and Arakan. He penned the chronical dynasties of Arakan and hinted the correlation among neighboring kingdoms around Arakan as well as portrayed the Hindu, Buddha and Muslim reigns.

Historian Abdul Karim in *The Rohingyas* (2000) demonstrated the Rohingyas' pristine history in Arakan that was enriched in every aspect of life. He clearly established the fact that Rohingyas are the progeny of Arabic and Persian settler who later was merged with immigrants of colonial era. Mohammed Ashraf Alam in *A Short Historical Background of Arakan* (1999) also emphasized the early historic Muslim contact in Arakan. Referring to the Mrauk-U Muslim dynasties, M A Tahir Ba Tha in *The Rohingyas and Kamans* (1963) claimed that there was no way to separate Rohingya Muslims from early established Muslim inhabitants like Arabic, Persian, Turks, Moors and Bengalese. Mohammed Yunus in *A History of Arakan: Past and Present* (1994), stated that the Rohingya Muslims origin goes back to Arabic and Persian, however, Maughs (Arakanese/Rakhine/Buddhist) went first from India and Southeast Asia. Azeem Ibrahim in *The Rohingyas: Inside Myanmar's Hidden Genocide* (2016) clearly interpreted the holocaust against Rohingya Muslim in Arakan. The book discloses Burmese authority's organized segregating and persecuting policies.

Khin Maung Yin in *Salience of Ethnicity among Burman Muslims: A Study of Identity Formation* (2005) noted that as a progeny the Rohingyas receive Islam from Arab Muslims who reached in Arakan in the eighth century.

Moshe Yegar in *The Muslims of Burma* (1972) traced the history of Muslims in Burma. He traced the evolution of Muslim community from early historical era depicting the struggle of the Muslim ethnic groups; however, he hinted that the Muslims had been stateless under Burmese authority. Aye Chan in *The Development of a Muslim Enclave in Arakan (Rakhine) State of Burma (Myanmar)* (2005) analyzed the dialect 'Rohingya' used first in 1950. He also claimed that the Rohingyas are progeny of Bengali Muslims who migrated to Arakan during British rule.

## 4. THE ROHINGYAS: A CASE FOR ETHNIC NATIONALISM

The data from interviews suggest that the Rohingyas believe Rakhine State has been the land of their ancestry since their ancestors have been living there for generations. Though they do not have concrete evidence to showcase they often refer to a thousand-year living history since the place was known as Arakan. As they strongly believe that their ancestors have been living in Rakhine State for centuries, they have developed a sense of belongingness with the land and the people who share same history, religion, culture, ethnic ancestry, heritage *etc*. Thus, as they socially evolved over thousand years, they have developed a sense of community identity, 'Rohingyas', which they strongly consider to be different from other Muslims living in Myanmar and Bangladesh.

The Rohingyas consider their identity to be firmly linked to 'the land' they live in, Arakan/Rakhine State. This has been central to their identity along with other elements already mentioned. Thus, the sense of belongingness which they have gradually developed over generations may be denoted as evolutionary development of a case of ethnic nationalism. Their sense of togetherness comes from the belief that their ancestors have been living there for centuries, the land belongs to them and they have an inherent right to live there.

In 1948, with the independence of Myanmar, a territorial nationalism was supposed to emerge in Myanmar, a nationalism in terms of the boundary of newly founded state which was supposed to be inclusive of all internal ethnic groups. Even though in the beginning it was more likely the case with political representation from all groups, however, gradually under the military rule it turned out to be exclusionary. Over the decades it failed to ensure representativeness of all groups.

In the case of Rohingyas, the new consciousness of unity is propagated and rejuvenated by the educated and politically conscious elite class referring to and based on the historical linkages to their ancestors in Rakhine State. The Rohingya diaspora living around the world shows the same sentiment. The Rohingyas thus emerged as a case for ethnic nationalism, a nationalism based on its ethnic focus as the territorial nationalism failed to meet their needs.

## 5. ARAKAN: A BRIEF DESCRIPTION AND ITS ANCIENT HISTORY

### 5.1. A Brief Description of Arakan

Arakan, the historical name, was replaced with 'Rakhine State' by the government of Myanmar in the 1990s (Bahar 2012). Arakan/Rakhine State refers to the extended coastal region of Burma/Myanmar, more specifically the western part of the country. It covers an area of almost 36,778.05 square kilometers (Census Report 2015: 17). Due to distinct geographical position surrounded by hilly areas, Arakan managed to survive as a separate, independent kingdom for long time until 1784 (Yunus 1994). Prior to that, there were two kingdoms in Arakan: South Arakan or Sandoway and North Arakan or Arakan proper; these two kingdoms were merged in the thirteenth century (Karim 2000). As per the Census Report 2014, the estimated demographic figure of Arakan, including 1,090,000 non-enumerated population who belongs Islamic religious faith, was about 3.18 million with two major ethnic groups 63.32 percent Buddhists and 34.18 percent Rohingya Muslims (Census Report 2016: 02–06).

Arakan is mountainous region on the eastern shore of the Bay of Bengal. In the north, it is bounded by India and in the East, the Yoma Mountains with 2,000 meters height form a natural fence between Arakan and rest of the Burma (Mohajan 2018: 28). As a coastal region, it has been easily accessible by sea routes (*Ibid*.). The River Naaf in the north and west Arakan is a common borderline with Bangladesh. In the north, it is also bordered by Chin State of Burma and Magway, and Bago. Ayeyarwady region is situated in the East.[1]

The word 'Arakan' is thought to originate from Arabic or Persian as both have same meaning (Mohajan 2018: 26). Phayre explained that the name might come from native name Rakhaing, from which the modern European form Arakan is derived (Phayre 1883: 42). Early Buddhist missionaries called Arakan as 'Rekkha Pura' (Alam 1999). Tripura Chronicle Rajmala mentioned as 'Roshang' (Mohajan 2018). The poets of Arakan particularly Alaol, Ainuddin, Abdul Ghani and others referred Arakan as Roshang, Roshanga and Roshango Des. Abdul Karim mentioned that Rohingyas believed the word Arakan is derived from the Arabic word all-Rekan or al-Rukn. In the sixteenth century, the Europeans used the name as Arakan (Karim 2000: 18–19).

## 5.2. The Ancient History of Arakan

According to Phayre before the recorded history, a son of the first King of Banaras was allocated to the historic land of Arakan (Phayre 1883: 42). According to anecdote, ten brothers of the Sandoway heir were killed, while a sister survived and escaped to the north of Arakan proper and later she married a Brahman. When the Brahman man came to the throne, Brahman (Hindu) dynasty was established. Phayre mentioned the King Marayo as a newborn baby who was found by the Chief of the Mro Tribe during his hunting in the jungle. Marayo married the daughter of Mro Chief (*Ibid*.: 43). After the marriage of Marayo, descendant of Brahman King, the capital city, Dhanyawadi, was founded.

In the upper Burma, there were a group of small Pyu city-states. The Pyu city-states were later absorbed in the Pagan Dynasty as the latter expanded and became an Empire (Phayre 1883). The Pagan Empire gradually evolved into modern day Burma/Myanmar.

### 5.2.1. Arakan: The Early Dynasties

In the early history, Arakan was an independent kingdom ruled by the Hindus, Buddhists, and Muslims. Dhanyawadi, the first recorded capital of Arakan Kingdom, was founded in 2666 BCE (Alam 1999) which existed until 327 CE. There is a record of 25 successive kings reigning in Dhanyawadi. Referring to an anecdote Singer (2008) indicated that initial wave of Hindu colonel took place on the first century by the sea voyagers who sought for commerce of gold, silver and so forth. Once they reached the shore of Arakan, they established Hindu Kingdom (Singer 2008).

The second phase of Indianisation of Arakan happened from the fourth to sixth century. The capital of the second dynasty was in the town of Waithali (Vesali). However, archeologist Aung Thaw pointed out that a Hindu dynasty was ruling the Vesali until the second century BCE (Singer 2008: 10). However, the majority argued that Vesali was established in 327 CE and in the reign of the second dynasty, Arakan reached its utmost classical period of culture, which lasted until 818 CE. During that period, Mohamuni Buddha image was casted in Arakan (Alam 1999).

### 5.2.2. Chandra (Hindu) Dynasties in Arakan

The Chandra dynasty is traced from 788 CE to 957 CE. In the early Christian era, Buddhism reached Arakan earlier than other parts of Burma. However, before the tenth century the original inhabitants of Arakan were solely Hindus (Tha 1963). Archeological findings indicate that Hindus predominated until the eighth century. Moreover, the Arakan was influenced by Brahmanism (Yunus 1994). Harvey further mentioned that Brahmanism is indicated by the word 'Chandra' that ends the name of every traditional king in 788–957 CE (Harvey 1925: 137).

Hindu kings ruled Arakan from the first to the tenth century (Alam 1999). According to M A Tahir Ba Tha, after Hindu civilization, the names of Dhanyawadi, Ramawadi, Maygawadi and Dwarawadi dynasties, given by Hindus, are thought to be Bengalese (Tha 1963).

## 6. EARLY INFILTRATION IN ARAKAN: MUSLIMS AND BUDDHISTS

The Buddhists and Muslims entered Arakan during the very early history of Arakan. The Muslims and Buddhists infiltration into Arakan proceeded in the following stages.

### 6.1. Early Muslims Infiltration

The first sign of the Muslim settlement was found in 788–810 CE. Documentary evidence shows that the early Arab had contact with Arakan and Bengal (Karim 2000: 24). According to the Arakanese traditional history, during the reign of Mahataing Sandaya, several Arab ships wrecked off the coast of Rambi Island (Ramree), the Muslim sailors somehow escaped and swam onto the shore. In the Arakanese history, they are identified as foreigners. The king provided them with shelter. Gradually the Arabians assimilated with the locals, and later the king allocated land for them to settle (*Ibid.*: 24–25).

Moshe Yegar argued that the Muslims seamen who first reached Burma in the ninth century were the ancestors of Rohingya (Yegar 1972: 2; Al-Mahmood 2016). The shipwrecked Arabian Muslims absorbed Burmese language and customs and gradually became the second largest ethnic group of Arakan (Karim 2000: 25–26). Karim emphasized that the Rohingyas had been staying in Arakan for more than a thousand years. Similarly, Alam pointed out that in the pre-Islamic times, peoples like Arabs, Moors, Pathans, Turks and Bengalese came to Arakan as traders and preachers; many of them settled there, assimilated with the local customs and they are the people who are currently known as ethnic Rohingya (Alam 1999).

### 6.2. Early Buddhists Infiltration

There was a serious tension between Buddhism and Brahmanised Hinduism during the eighth century in India (Yunus 1994). In Magadha, old Bihar, chauvinist Hindus, and rival Mahayana Buddhist that compelled Hinayana Buddhists to flee eastward and they took harbor in Vesali dynasty of King Chandra (*Ibid.*). That was the first recorded large infiltration of Buddhists into Arakan. Since then they have been called Maughs, the word assumed to be derived from Magadhi (*Ibid.*). During Mongolians invasion in 957 CE, Indian Buddhists, and Hindus of Arakan merged with Mongolians or Tibeto-Burmese; however, Rakhine Maughs claim to be more Aryan than Mongolians (Tha 1963). After the tenth century, the kingdom professed Buddhism. However, before the arrival of Muslims the people of Arakan were the local Hindus and Buddhists who came from India (Harvey 1925: 137–138).

### 7. ARAKAN AND THE NEIGHBORS: HISTORY OF INVASIONS AND COUNTER INVASIONS

The history of Arakan developed through a complicated process marked by frequent accounts of invasions and counter-invasions over thousands of years. There are numerous historic records of invasions and counter invasions between and among Arakan and its neighboring political entities. Over thousands of years, many neighboring dynasties attacked Arakan. Arakan also raided many neighbors such as Chattogram and Pegu. Through these invasions and counter-invasions, numerous mass migrations took place and through this the ethnic groups and composition of the population formed in Burma and in Arakan.

**Arakan's Attack on Chattogram: The First Phase.** The historic record denotes that the Arakan King, Tsu-La-Taing-Tsandra (951–957 CE) invaded Chattogram first in 953 CE (Ali 1967). The King constructed a monument called Tsit-ta-gung (*i.e.*, No war will be there) for commemoration. The name of Chattogram is thought to be derived from that inscription (Ali 1967; Yunus 1994).

**The Pagan Kingdom's/Burmese Attack on Arakan: The First Phase.** The great King of Pagan Anoarahta invaded Arakan in 1044, desiring to hold the famous image of Buddha (Phayre 1883: 46). Meantime, under Pagan reign (1044–1287), the Burmese established their imperialism over the south Arakan. In addition, the Buddhism became modified from Mahayanist to Hinayanist as like Burma during the eleventh and twelfth centuries (Tha 1963). Harvey pointed that in the middle of the twelfth century the famous Mahamuni image was unavailable (Harvey 1925: 137–138).

**Bengal and Turkish Invasion on Arakan.** The Turkish Armies invaded Burma in 1277 under command of Nasser Uddin. Yegar interpreted that during the occupation another influx of Muslims from China occurred (Yegar 1972: 3). Finally, in 1285–1287, they attacked and destroyed Pegu. When the Europeans came into Arakan in the fifteenth, sixteenth and seventeenth centuries, they described the coastal cities of Burma as Muslim colonies (*Ibid.*: 3–4).

During the reign of Minhti (1279–1374), Arakan experienced a devastating sea attack from Bengal. The King successfully used some ploys to defeat invaders (Harvey 1925: 138).

**The Mongol Invasion on Burma.** The Mongols undertook a series of invasions on Pagan dynasty as the King Narathihapate denied demand of tribute by the Mongols. The Mongol undertook three invasions on Burma from 1275 to 1285. As the Mongols won the war and formed a new kingdom resulting in the formation of almost new population base, thus the Burmese race was considered mongoloid largely (Hall 1950).

**Burmese Attack on Arakan: The Second Phase.** In the early fifteenth century, Arakanese King Naramekhla, also known as Min-Saw-Mon invaded some area of Burma, however, vanquished quickly, and belligerent turmoil led to reverse attack on Arakan. The Burmese took possession of Launggyet, the capital town and the King Min-Saw-Mon was expatriated from his kingdom in 1406. Fleeing into Bengal, Min Saw Mon got asylum and resided in Gour, the capital of Bengal, for twenty-four years (Phayre 1883: 77).

**Restoration of Arakanese Kingdom and Muslims Infiltration.** The next phase of Muslim inflow in Arakan took place in the fifteenth century and the Muslims were brought into Arakan by the King who got back his throne with help of the king of Bengal, Nazir Shah, widely known as Sultan Jalaluddin Mohammed Shah (Karim 2000: 26; Islam 2017; Tha 1963).

In 1420, the King of Bengal, Sultan Jalaluddin Mohammed Shah experienced Delhi's invasion. During that time, the exiled King Min-Saw-Mon helped King of Bengal to set ploys to trap the enemies (Akhanda 2013: 38). As the King benefited, he determined to assist Min-Saw-Mon by restoring his kingdom. However, the first attempt failed in betrayal of his general, U-lu-Kheng also known as Wali Khan (Karim 2000: 27). Then the King, appointed his ministers, Dampasu, Razamani, Setta Khan headed by general Sandi Khan who restored Min-Saw-Mon to the throne in 1430 (Phayre 1883: 78; Tha 1963). As gratitude, the Buddhist king, Min Saw Mon sent tribute to the King of Bengal, and adopted Muslim titles called Sulaiman Shah. He also issued coins, bearing kalima, Muslim inscriptions etc. Many Muslims settled in Arakan at that time (Phayre 1883: 78; Karim 2000: 27–28).

While returning to throne, King Min-Saw-Mon moved the capital from Launggyet to Mrauk-U which was known as Roshang (Rohang) to Bengali. From 1433, Mrauk-U remained the capital of Arakan for the next four centuries (Harvey 1925: 139; Karim 2000: 29–30). The Muslim soldiers, who fought to restore Min-Saw-Mon, did not return to Bengal rather got appointed as high-ranking official of the royal court by the many kings of Arakan (Akhanda 2013: 42). After the death of Sultan Jalaluddin Mohammad Shah and Min-Saw-Mon the Arakan-Bengal relation, remain unchanged (Alam 1999: 9–15).

During the Mrauk-U Regime, the European and Arab traders visited the coastal ports, with the predominate presence of Portuguese as mercenaries and pirates. They were engaged with local community and as a result the number of Muslims increased gradually there.

**Arakan's Attack on Chattogram: The Second Phase.** The successor King of Mrauk-U, Narameikhla known as Ali Khan (1433–59) withdrew the tribute and annexed Sandoway and Ramu. The successive king Basawpyu known as Kalima Shah (1459–82) occupied Chattogram in 1459, and it remined under Arakanese control for about two centuries, until the Mughals occupation in 1666 and that was the second attack of Arakan in Chattogram (Harvey 1925: 140; Farzana 2017: 43).

The Arakan King, Min Yaza invaded Chattogram with 4,000 soldiers and a naval fleet in 1517–18. The Mughal governor fled to Sonargaon. The Arakanese occupied Sandwip and Hatia with headquarters at Lakhipur. However, under great king Minbin, Arakan steadily established authority over Chattogram by the mid-sixteenth century and controlled it during the most part of the sixteenth century.

After 1532, during the King Minbin as Zabuk Shah (1531–53), who was the greatest King in the history of Arakan and founder of the 'Arakanese Empire,' the Feringhi or Portuguese inhabited on the coast of Arakan. The Portuguese arrived in Arakan in 1517 and King appointed them as military officers. During the reign, Arakan Empire became enriched with high-powered weapons (Yunus 1994). In the year 1538, Bengal faced two successive incursions from Afghan and Mughals that resulted King Minbin occupation of eastern Bengal including Chattogram (*Ibid.*). According to Harvey, since the middle of sixteenth century many Portuguese settled in Chattogram, making it a thriving port (Harvey 1925: 141).

**Bengal's conquest on Arakan: The Second Phase.** The Bengal Sultan's second conquest of Arakan took place in 1554 and it was led by Shamsuddin Abu Muzaffar Mohammad Shah (Yunus 1994). Minbin's son, Mingphalaung bearing the name Sikandar Shah (1571–93) brought back the throne as well expanded the occupation up to Decca and all parts of Chattogram, Noakhali as well as Tippera. Later, in the reign of Minyazagyi, also known as Selim Shah (1593–1612), a son of Sikandar Shah, extended Arakan from Decca and the Sundarbans to Moulmein with 1,000 miles in length and 150–200 miles in depth (Alam 1999). Selim Shah also massacred 600 Portuguese settlers to prevent incursion in 1607. However, Portuguese pirates returned with weaponized followers at Sandwip island and settled there, ousting Afghan pirates in 1609 (Harvey 1925: 141–142).[2]

King Minhkamaung bearing name Husein Shah (1612–22), weakened the Portuguese powerbase at Sandwip. The initial advance on Sandwip was revoked due to Tippera invasion. In the consequence, Arakan met a counterattack by Portuguese in 1615. Eventually, in 1617, Husein Shah beat them off with the help of Dutch. The Portuguese were ceased and became the servants (Harvey 1925: 142).

The Portuguese formed an alliance with Arakan Maughs. Afterwards Chattogram became a breeding ground of pirates who looted and plundered entire lower Bengal. From 1621 to 1624, the Arakanese Maughs and the Portuguese pirates captured 42,000 people including Muslims and Hindus from Bengal and sold them as slaves (Alam 1999). Throughout the seventeenth century, the Arakanese Maughs and Portuguese continued slave-trade at the ports of Arakan and India. Harvey noted that in one month, February (1727), the Maughs and Portuguese carried off 1800 captives from the southern Bengal and sold as slaves in Arakan. This gives a rough estimation about the total number of people sold as slaves during this time. The Maughs and Portuguese continued slave trade for a long-time turning Bengal into a desolate place (Harvey 1925: 143–144). Throughout the eighteenth century, their slave trade and brutalities continued.

**Arakan's Attack on Pegu.** When Arakan established authority over Chattogram and parts of Bengal, it launched an attack on Pegu in 1599. The King Meng Razagvi employed flotilla from Chattogram and the Ganges delta (Ali 1967).

**Mughal Prince Shah Shuja's Shelter in Arakan.** The Mughal Prince Shah Shuja took shelter in Mrauk-U in 1660. This caused many Muslims' migration to Mrauk-U (Harvey 1925: 146). When Arakan King Sanda Thudamma killed Shah Shuja, he incorporated Shuja's companion Muslim soldiers into his elite guard, 'Kaman' (Karim 2000: 48). Later, in 1710, the members of 'Kaman' were exiled to Ramree Island (Phayre 1883: 178–179; Harvey 1925: 148). Muslims were recognized as indigenous ethnic group in Arakan and many held high positions in the society as lawyer, teachers, doctors, civil servants, and other professionals (International Crisis Group 2014).

**Burmese Invasion on Arakan: The Third Phase.** Burma annexed Arakan in 1784. The Burmese King Bodawpaya undertook the third attack on Arakan in 1784–85. It was a massive assault with twenty thousand soldiers, two thousand five hundred horses and over two hundred elephants causing a huge destruction in Mrauk-U (Phayre 1883: 213). The Arakan King's family was forced to move to Upper Burma and tens of thousands of Arakanese Muslims were taken as slaves, more than 35,000 fled into Chattogram and 40,000 men were killed by the Burmese (Chan 2005). In 1785, first exodus of Rohingya Muslims into Bengal was recorded. In 1790, the British-Indian administration sent diplomat Hiram Cox to Bengal territory to assist refugees and he established Cox Bazar (Al-Mahmood 2016).

**The British Annexation of Arakan and Burma.** The British Empire annexed Arakan in 1824. As the expansion of Burmese Empire bordering Bengal (*e.g.*, Dhaka, Kolkata) seemed threatening to the British, so, they annexed Arakan in 1824 from the Burmese. The British waged two more wars in 1852 and 1885 and annexed the whole Burma (Steinberg 2010: 27) to British Empire, or the British-India (Yegar 1972: 29; Karim 2000: 112–116).

The first civil ruler of Arakan, Mr. Robertson (Karim 2000: 112) in an assessment report emphasized the prospect of massive rice and other agro-commodities production. He suggested importation of cultivation-skilled peasants from neighboring zone, Chattogram and India. He assessed population composition of Arakan before the British conquest as 60 thousand Maughs, 30 thousand Muslims and 10 thousand Burmese (*Ibid.*: 115). The main Muslim groups include Thambaikkya, Zerbadi, Kamanchi and predominant Rohingyas (Yegar 1972: 29). As Burma became a province of the British-Indian Empire, Indians could enter Burma easily.

With concentration on increasing rice production over extensive farmlands, more and more people went to Arakan from Bengal. There was also seasonal flow of laborers from India. The colonial administrators further encouraged labor migration into the tea and rubber plantations, and among migrants there were mostly Muslims and some Hindus (Sarkar 2018). Construction of a railway from Buthidaung to Maungdaw facilitated more transportation of seasonal Bengali laborers (Yegar 1972: 29, 48; Sarkar 2018).

The British preferred Indians for high administrative positions. In the beginning, they could get positions of government clerks. Later,

Indians went to Burma as teachers, engineers, businesspersons, bankers and so on (Yegar 1972: 30).

Skilled and talented Indians took urban professions, thus concentrating mostly in the cities, but considerable number also settled in villages. In Rangoon and in most of Burma's towns and villages, the traders and shopkeepers were usually Indians, mainly Muslims. Indian migrants made sure that other Indians joined them so by the early twentieth century, their number reached one million with more than half of them being Muslims (Yegar 1972: 30–31). Aye Chan argued that people were brought from Bengal and British administrators granted them land (Chan 2005).

## 8. ARAKAN UNDER THE BRITISH CROWN COLONY

Arakan went under 'Crown Colony' administered by the British in 1937. Under the Act of Government of Burma in 1935, Burma became a distinct Crown Colony in 1937 (Sarkar 2018). The British stimulated labor flow, hundreds and thousands of people from India went to Arakan for agricultural needs (Farzana 2017: 44–45).

During WW-II, in 1942, the Japanese air force attacked Sittwe (Akyab) resulting withdrawal of British administration from Burma to India. The Burmese assisted the Japanese occupation forces aspiring for their independence (Farzana 2017: 17). Since the Muslims community was pro-British, the Japanese conquest shifted ethnic balance towards the Burmese. This resulted in rivalry among Burmese and other ethnic groups causing communal violence. Both Buddhist and Muslim communities formed armed units and attacked each other. Human Rights Watch reported that 22,000 Rohingyas crossed the border into Bengal during that time (HRW 2000; International Crisis Group 2014). The Japanese undertook multiple assaults massacring the Rohingyas for their pro-British stance and destroyed 307 villages, 100,000 Rohingyas died and about 80,000 fled from the area. Ibrahim pointed that the ethnic violence in 1942 caused segregation of Muslims to the north and Buddhists to the south (Ibrahim 2016). Rakhine remained under the Japanese control until the end of war.

'Thirty Comrades'[3] including General Aung San formed the Burmese Independence Army (BIA) to assist the Japanese against the British allies. The BIA also initiated to negotiate with the radical Muslim leaders at Maungdaw but failed. Meantime, the Arakanese Muslims started dreaming of a separate independent region (Ibrahim 2016). The Volunteer Force and British-Indian troops gradually weakened Japanese through series of random attacks. Aung San started to negotiate with the British and their independence movement subsequently switched sides and took part in the attack against Japanese from March 27, 1945. The British force occupied Sittwe (Akyab), the capital city of Arakan in December 1944 and arrested Burmese guerrilla leaders (Chan 2005). Some of the displaced Muslims returned with the British. However, the British did not keep their promise of creating a place for the Muslims. In the post-war period, distrust erupted between Buddhist and Bengali Muslims.

The Muslims formed Muslim Liberation Organization (MLO, later renamed Mujahid Party) under the leadership of Zaffar Kawal and Abdul Husein in March 1946. The MLO members collected arms and became Rohingya's own army. In 1947, during creation of India and Pakistan, the Rohingya leaders wanted incorporation of northern Arakan with East Pakistan but failed (Farzana 2017). Moreover, after some Arakanese Muslims filed a petition in the Constituent Assembly in Rangoon after independence for integration of Maungdaw and Buthidaung into East Pakistan. The Burmese authorities considered the Arakanese Muslims as hostile to the new regime and later labeled them as outsiders (Ibrahim 2016) whereas the regime considered the Buddhists as part of the new state.

## 9. THE INDEPENDENT BURMA/MYANMAR

In 1948, Burma became independent and Arakan became part of the independent country. The Rohingyas got recognition to be part of it.

### 9.1. Arakan and the Rohingyas under Burma

Ethnic conflicts erupted soon after establishment of independence. One of the major reasons of Mujahid rebellion was denying Muslims to resettle in their villages who fled from Arakan during Japanese invasion (Yegar 1972: 66–67). Moreover, the Buddhist officials positioned in the colonial administrator seats, imposed restrictions on the movement of Muslims from northern Rakhine to Sittwe. After the War, Muslims who went to East Pakistan were not permitted to return (Chan 2005; International Crisis Group 2014).

The Mujahid Party sent a letter to the Burmese government stating demands of a national home for Muslims, recognition of Burmese nationality, resettlement of the refugees from the Mrauk-U Townships, general amnesty for the Mujahid Party members and recognition of the party as legitimate political organization (Chan 2005). As state authority ignored the Rohingyas demands, in 1948, the Mujahidin destroyed all the Arakanese Buddhist villages in the northern part of Maungdaw, Ngapru-chaung and nearby villages in Maungdaw

Township. In all Arakan Muslim Conference in 1951, Mujahid leaders demanded the balance of power between the Muslims and the Maughs (Buddhist) (Chan 2005) and that North Arakan should be given the same status as other states (specified for Muslims of Arakan) like, the Shan State, the Karenni State, the Chin Hills, and the Kachin Zone with its own Militia, Police and Security Forces under the General Command of the Union of Burma (*Ibid.*). In the early 1950s, Mujahid's armed attacks caused large influxes into East Pakistan from Maungdaw, Buthidaung and Rathedaung. Pakistan government warned Burmese Authority about how they treat Muslims in Arakan (HRW 2000). However, later in a diplomatic turn, the Burmese Prime Minister U Nu approached Pakistan to stop supporting them by all means (Sarkar 2018).

The Mujahid leaders identified themselves as the Muslims of Arakan. Mr. Abdul Gaffar, the MP from Buthidaung mentioned the word 'Rohingya' for first time in his article in 1951 (Chan 2005). Ibrahim argued that the concept of the 'Rohingyas' is a recent construction is one that the regime favors (Ibrahim 2016). Tonkin argued that historically Rohingya was never used as an ethnic designation rather it was fashioned after the WW-II to define Muslims residing in the north of Rakhine State (*Ibid.*).

In the independent Burma, four Muslim leaders from Buthidaung and Maungdaw townships became member of legislature in post-1948 election (Chan 2005). During the 1960 elections campaign, the Burmese Prime Minister U Nu promised safe statehood for the Arakanese Maughs and when he planned to form a state for Maughs, the Muslim representatives vetoed and demanded for the Rohingya Muslims State as well (*Ibid.*).

General U Nu proclaimed Rohingyas to be equal to other ethnic groups in Arakan on several occasions. The Burmese government (1948–1962) recognized the Rohingyas as citizens (Poling 2014). Rohingyas were included in 1961 census as an ethnic group in Arakan (Ibrahim 2016) and got National Identification Card with access to all benefits of citizenship (Zarni 2013).

### 9.2. Arakan and Rohingyas under Military Rule of Independent Burma

Arakan went under rule of the military regime after 1962 and the situation started getting worse (Ibrahim 2016). General Ne Win and his Burma Socialist Programme Party began dissolving Rohingyas social-political organizations. The government privatized assets of Indian, Chinese, and Pakistani entrepreneurs resulting about 100,000 Indians and 12,000 Pakistani people leaving Burma (Chan 2005). In 1963, the Rohingya Independence Force (RIF) was formed with same philosophy of the Mujahidins to seek separation of Arakan (Yegar 2002: 53). The RIF was succeeded by the Rohingya Patriotic Front (RPF) in 1974 (Selth 2010).

The reformation of Constitution in 1974 dropped the status of the Rohingyas. In 1977, the Burmese military ruler conducted operation Naga Min causing further sufferings for the Rohingyas (HRW 2000). Military authority forcibly pushed out more than a quarter million of Rohingya Muslims in 1978 who fled to Bangladesh (Yegar 2002: 56).

The Citizenship Law of 1982 recognized 135 ethnic groups but excluded the Rohingyas (Riaz 2018). The law legitimated citizenship for them who were in Burma prior to 1823, before the First Anglo-Burmese War in 1824 and thus deprived Rohingyas from all state facilities (Ibrahim 2016). During the anti-Muslim riots in 1983, the Buddhist Maughs burned down Karen Muslim's quarters and mosque in Moulmein and Martaban city. Consequently, a Muslim organization was formed in 1984 known as Kawthoolei Muslim Liberation Force (KMLF) and apparently 200 members of the organization were weaponized and equipped by Karen insurgents. In addition, the members of KMLF took part in violence against Burmese army near Thailand border (Yegar 2002: 60).

The new military regime seized power in 1988 and renamed the country from 'Burma' to 'Myanmar' in 1989 to promote exclusionary nationalist agenda (Mohajan 2018: 28; BBC 2017). In the 1990s, the name Arakan was replaced with 'Rakhine State' (Ludden 2019; McKenna 2017). The junta began settling Buddhists in Muslims area, Northern Rakhine, and they engaged in murder, rape, and rupturing mosques (Yegar 2002: 63). In 1991–92, another massive exodus from Rakhine to Bangladesh took place and nearly a quarter million of Rohingyas fled to Bangladesh (HRW 2000). Prior to that, while pursuing Rohingya rebels, the Myanmar military attacked the Bangladesh army camp and killed a soldier; left some injured (Yegar 2002: 63). With the UN agency reconciliation process from 1994 to 1996 some 200,000 Rohingyas were repatriated to Myanmar and more than 200,000 stayed in Bangladesh (*Ibid*.: 66).[4]

Between 2008 and 2012, Tatmadaw destroyed many mosques and built many Pagodas in Rakhine. Buddhists got upper hand and used Rohingyas as forced labor and imposed restriction over traveling, schooling, and other daily activities. Some of them even accepted Buddhism for survival (Ibrahim 2016).

The 2012 violence forcibly displaced 140,000 Muslims including Rohingya, Rakhine, Kaman and Maramagyi, 100,000 Rohingyas were captured in internal refugee camps in Rakhine and about 20,000 fled to Thailand and Malaysia (Ibrahim 2016). According to UN report, 8,614 houses were destroyed causing 192 deaths, 265 were seriously wounded. In alliance with ethnic Rakhine the security forces looted and destroyed Muslim's shops (United Nations 2018: 150–151). Previously, in 2014 national census dialect 'Rohingya' was replaced with 'Bengali' and in 2015, they were denied right to vote in general election (Riaz 2018).

### 9.3. Contemporary Scenario of the Rakhine State and the Rohingyas

The UN reports posit that on 25 August 2017, Arakan Rohingya Salvation Army (ARSA) launched coordinated attacks on military base and up to 30 security force outposts across northern Rakhine State. Twelve security personnel were killed; however, the villagers had light homemade weapons, for instance, sticks, knives, swords, slingsshoots, and some homemade explosives (United Nations 2018: 180, 242). The counter response of Tatmadaw was sheer ethnic cleansing. The operation encompassed hundreds of villages across Maungdaw, Buthidaung and Rathedaung Townships. The UN said 40 per cent of the villages of the townships were totally or partially destroyed and that resulted in massive flow of Rohingyas to Bangladesh, numbered over 725,000 Rohingya Muslims including some Hindus (*Ibid.*: 180). The UN officials considered the Burmese authority's treatment to Rohingya as a 'textbook example' of ethnic cleansing (Safi 2017) and the UN Secretary General, Antonio Guterres, denounce the operation as ethnic cleansing (Besheer 2017). Ali Riaz noted that exclusionary nationalism was an important element of Rohingya expulsion (Riaz 2018).

### 10. THE SOCIAL EVOLUTION OF THE ROHINGYAS

Since ancient times, the social evolution of the Rohingyas has been entwined with social evolution of the Muslims in Rakhine State. The Muslims community developed in Arakan starting from the late 700 CE when Arakan encountered the Arabs. During Mahataing Sandaya's reign (788–810 CE), some Arabian Muslim sailors were allowed to live in Arakan after their ships wrecked off the Ramree coast. As depicted above, though the Muslims initially had marginal role in the society, they gradually became influential and important in the society. The Arabian Muslims who got there from a wrecked ship, they gradually became the mainstream in Arakan and in Arakanese society. They are considered as the ancestors of Rohingya (Karim 2000: 24–25). From the ninth to twelfth centuries they became important since the Arabian merchants expanded trade along the South and Southeast Asia coastal regions. During this time, more and more Arabian and Persian traders came to settle in lower Burma and Arakan and since twelfth century, Muslims had significant presence in northern Arakan (Yegar 1972: 2–3). The Arabian and Persian Muslims assimilated with the locals and absorbed Arakanese culture and language without relinquishing their ancestor's religious belief.

The King Sawlu, educated by Arabian Muslims, like his father king Anoarahta sheltered the Muslims in Arakan. When King Sawlu ascended to the throne, he appointed his teacher's son Yaman Khan as governor of the city Pegu. Later Yaman Khan became disloyal to the King. Once Yaman killed the King Sawlu and ascended the throne. It was the first time when Muslims gained the throne and royal power (Yegar 1972: 3). With the Muslims in royal power and his treatment of the Muslims, the social status of the Muslims was uplifted in Arakan society.

## 10.1. Social Evolution of the Rohingyas: Early Arakan Era

The early Arakan era started with its inception as an independent kingdom, distinct from Burma and Bengal till British conquered Arakan. Prior to this era, Burma invaded Arakan in 1406 and the King of Arakan, Min-Saw-Mon, fled to Bengal. In 1430 Min-Saw-Mon regained the throne with the help of Bengal Sultan Jalal-ud-din Muhammad Shah. The Muslim soldiers and their escorts, who went to Arakan to fight against Burma, settled there to support further Min-Saw-Mon. As gratitude, the king adopted Muslim name 'Sulaiman Shah', issued coins with Muslim inscriptions, and appointed Muslims on the top of divan (Phayre 1883: 78). The new Muslim population married and settled there.

Through the stalwart efforts of the new Muslim community members Arakan reached peak of its flourishment. In the Bay of Bengal region, capital of Arakan Kingdom, Mrauk-U, became a significant regional trade hub. Integrated with assiduous Muslim newcomers, the Kingdom built up a powerful naval force and established unprecedented business hub in the coastline of the Bay of Bengal. It was the high time for Muslims community in Arakan (Alam 1999: 14).

The incorporation of the accompanying soldiers of the Mughal Prince Shah Shuja, when he took refuge in Arakan (1660), into the elite palace guard as a special unit (*i.e.*, Kaman), was an important event in the evolution of the Muslims in Arakan (Karim 2000: 48).

Thus, from fifteenth to seventeenth centuries, Muslims' position became stronger and importance increased in Mrauk-U. Also, a good understanding between Akaran and Bengal contributed better positioning of the Muslims in Arakan as many of them got place in the upper strata of the society through various professions, for instance, lawyers, teachers, doctors, civil servants etc. (International Crisis Group 2014).

The Burmese occupation of Arakan in 1784–1785 disrupted both Muslim and Maughs community. The usurpers thereupon took Arakanese pretty women and girls without any consideration of their families moreover killed about 40,000 men in a day (Chan 2005). In addition, many Muslims were forced to move to Upper Burma and approximately 200,000 Arakanese (Rakhine) fled to Chattogram (Bengal). Later, King Bodawpaya included Muslims in military force as a new unit known as 'Myedu' (International Crisis Group 2014). During the four decades of tyrannies, Burmese authority persecuted Muslims and Arakanese in many ways as they massacred Arakanese who were unable to pay tax, even infants were not excluded from poll-tax. Around 3,000 people were sent to forced labor in the Meiktila lake construction who never came back, besides around 6,000 more died of diseases after dragging away to serve against Chiengmai (Harvey 1925: 280).

### 10.2. Social Evolution of the Rohingyas: The Colonial Era

During the Colonial Era (1824–1948) the British annexation of Arakan in 1824 (later Burma in 1885) had a significant impact on social and economic status of Rohingyas in Arakan. During the British occupation of Arakan there was about 30,000 Rohingya Muslims among 100,000 total populations. Observing the potentiality of growing agricultural commodities on fallow land, British authority brought people from neighboring India and Bengal into Burma. During that time, many Muslims and Hindus migrated to Arakan which gave an impetus to Rohingyas to produce a powerful influence over society (Karim 2000: 112–115).

Migrated Muslims formed family by marrying local Buddhist and Muslim women. Burmese women married Muslims and converted to their husbands' religion. The Muslims tremendously contributed to the increase of farmland and rice production. Whereas in 1845 the farmland of Burma was 354,000 acres, in 1930 the area grew to 12,370,000 acres (Yegar 1972: 29–33). In the 1910s, in the Maungdaw Township alone, there were fifteen Bengali Zamindars who brought thousands of Chit-tagonian tenants and established agricultural Muslim communities and built mosques with affiliated Islamic schools. However, in the British records all these villages occupied by the Bengalese continued to be called by Arakanese names (Chan 2005).

As Muslims settled down, they built Mosques and Madrasas which were open for their children to learn Quran, Arabic, and Urdu. The religious orthodox Muslims became Burmese in all other aspects except religion (Yegar 1972: 30–31, 40). Moreover, more Indian Muslims were brought into Arakan. They went there and occupied the top administrative positions from public service to military (Akhanda 2013: 54). As the British favored people from different ethnic and religious groups, they were able to reach various top levels of colonial services (Farzana 2017: 44). So, Muslims' social status was high in the society. As time went on, Muslims became important factor for social and economic leadership; moreover, they also became a catalyst of development in Rangoon and some other towns. Furthermore, their social and economic status dominated other ethnic groups in Burma.

In the early twentieth century, Muslims became cognizant of social activities as they formed various social organizations. In 1908, Bohras Soorti Mohamedan Association was formed in Mumbai and most of its members were wealthy merchants. To build a unique socio-economic and cultural life for the Burmese Muslim community, a Muslim Society named 'U Bah Oh' was founded in Yangon in 1909 and was funded by Muslim businessman (Akhanda 2013: 55) which later raised its voice for Muslims rights in Burma. At the same time, the Rangoon Meiman Jamat was established. A sporting club, the Islamic Fraternal Society established in 1909 marked obvious socio-cultural consciousness. A young group of Shia Muslims established Persian Association in Rangoon in 1909 to address social and religious issues. In 1930, it was reformed into Iran Youth League and in 1935 the name of the association was changed to Iran Club when Persia was changed to Iran; this was one of the oldest social organizations in Burma (Yegar 1972: 40–41).

In 1912 the Malabar Muslims established Cholia Muslim Association. Their community members were scattered throughout Burma conducting small businesses. In 1946, it was renamed into the All-Burma Tamil Muslim Association. Another association called Malabar Muslim Association was formed in 1918. Bengal Muslims had many associations, such as, the Decca Club, the Chittagong Association, and the Bengal Association. All these associations merged into All-Burma Pakistan Association after Pakistan was established in 1947. Similarly,

Qadiyani and Ahmadis believers also established association at the beginning of the century (Yegar 1972: 41).

Several Muslim associations were registered and recognized by government as scientific, literary, or charitable organizations. Their activities were almost the same, consciously avoiding political affairs, concentrated on cultural, societal, religious, educational spheres and community's welfare. Moreover, occasionally they also cooperated with non-Muslims associations on the issues of general welfare. As Yegar noticed the actual achievements of most these associations was much more modest than their declaration as they devoted themselves to development of religious, social, cultural, and educational status. Many schools, libraries, mosques, hospitals, and social centers were established by these associations and acted as an arbiter to mitigate inter-community conflicts; moreover, they announced scholarships for poor and brilliant students from all religious groups. Thus, Shia association was active in communal activities, while the Iran Club regularly arranged sports competition. The All-Pakistan Muslims Association came into public attention after World War II when they served as an umbrella organization for various Bengali Muslims societies (*Ibid.*: 42–44).

All Muslim groups lived in harmony and practiced their rituals without any conflict. This marks a tremendous progress of the societal evolution. The Muharram celebration was held with excessive enthusiasm. Shia community was permitted to flagellate themselves in Burma; however, to avoid any accidents, emergency ambulances were prepared by police. Furthermore, the Dawoodi-Bhora community took part in parade. The Dawoodi-Bhora had no community association and they were under the leadership of a single, Imam, who came from India (*Ibid.*: 45).

Beside these communal associations, each community maintained close ties with home district in India and Bengal. Consequently, in 1909 Indian Muslims formed All-India Muslims League in Burma, however, later it was transferred to Khilafat movement in 1918. All Muslims community also established several general organizations and institutions for Burmese Muslims welfare, with particular emphasis on youths' education (*Ibid.*: 47–48).

During that time in Burma, most of the mosques were used as schools. Besides, many schools were established in Rangoon by Muslims such as the first secondary school and Madrasa Mohammedia Randeria High School (started in 1867), one of the oldest educational institutions in Burma. When government included these institutions in formal education curriculum, children from Hindu and Buddhists community along with Muslims, began attending schools from 1927. Islamia School was another significant Muslims secondary school (established in 1886) whose board of directors included representatives from all Muslims sects. Prior to World War II, there were forty-one Muslim schools in Rangoon and eleven of them were private. All these school published monthly and annual journal in Urdu and English. Furthermore, several orphanages were built in different parts of Burma. During the British period five orphanages were established in Rangoon and some of them were annexed to the Schools and Madrasas (Yegar 1972: 50–51).

In addition to children education, the Muslim community began to arrange annual educational conference named All-Burma Muslims Educational Conference since 1905. Even scholars and university students began to engage in teaching to uplift the standard of Muslim education. As a result, the unity and status of Muslims was strengthened. In 1921, the Muslim students of the University of Rangoon formed the Rangoon University Muslim Students' Association. They established a library with Muslim donors' support and published annual journal on religion, politics, literature and history and organized lecture and lessons on religion. Due to internal conflict, the organization split into two parts and another organization, the Muslim Students' Society was formed in 1923. They began to work for welfare of Muslim society. They would also publish journal named 'The Cry' and distributed it for almost free of cost; moreover, another Muslim Students' Society established in Maymyo, in 1950 with the similar constitution and responsibilities (*Ibid.*: 52–55).

By the mid-twentieth century a new educated and politically conscious younger generation superseded the older ones. Before the beginning of the Second World War a political party, Jami-a-tul Ulema-e Islam was founded under the guidance of the Islamic scholars of which Islam was the ideological basis (Chan 2005).

During the World War II, when Japanese forces invaded Burma, as Muslims supported the British whereas Burmese supported the Japanese, the Burmese Independent Army along with Japanese force killed 100,000 Muslims and forced 500,000 to flee. As Japanese carried out atrocities on Muslim community, most of Muslims organization became inactive. However, Volunteer Force Muslims fought against Japanese and Burmese. By that time, Muslims community became more aware of their political rights as a result they went on demanding a separated Muslim area in northern Arakan. The Muslim Liberation Organization urged British and Pakistan to incorporate northern Arakan into the newly created East Pakistan (Chan 2005).

After the War, the Burmese nationalist leaders founded 'Anti-Fascist Peoples Freedom League' (AFPFL) in 1945. To bring the Muslims into the wave of that Nationalist movement the Muslim leader Shiraji Abdur Razzak established 'Burma Muslim Congress' (BMC) in 1945. BMC became a part of AFPFL later on and contributed to nationalist movement. Another Muslims' organization 'General Council of Burma Muslim Associations' (GCBMA) demanded for distinct rights for Muslims in Burma Constitution; however, AFPFL leader Aung San and Abdur Razzak pursued them not to be excited for regional or communal rights prior to independence (Akhanda 2013: 57–58). After the assassination of Aung San and Abdur Razzak, GCBMA made Muslim leaders united towards Muslims rights.

### 10.3. Social Evolution of the Rohingyas: The Post-Colonial Era

In the beginning of the post-Colonial Era (1948 onwards), Muslims began engaging in politics and Mujahid Party (MLO) initiated to negotiate Muslims' rights with the government. Myanmar enjoyed democracy from 1948 to 1962. In the 1960s election, four Muslim leaders became members of Parliament and they raised voice for Muslims rights (Chan 2005). In the meantime, the name 'Rohingya' received much attention and popularity. In 1961 the Rangoon University Muslim Students' Association protested vigorously against the proposal of introducing Buddhism as state religion in Myanmar (Yegar 1972: 86). They also formed Rohingya Independence Force (RIF) in 1963 to protect their rights. However, in accordance with the 1974 Constitution, the military government established Buddhism as the state religion and disbanded Rohingya's social and political organizations consequently all the organizations went underground (Farzana 2017: 48).

Things started getting worse since the 1960s. The Rohingyas were denied formal citizenship in 1982; hence, some of them allowed vote in the 1990 general election. Rohingyas' political party 'National Democratic Party for Human Rights' was established in 1989 and won four seats but after the election the military authority banned the party (Farzana 2017: 53). After the 1990 general election, by no means Rohingya could participate in national of local government election. The 'National Democratic Party for Human Rights' registered as Democracy and Human Rights Party and contested in the 2015 general election. Only three out of eighteen candidates could participate but after a scrupulous check and cross-check the Tatmadaw authority made it certain that none of the Rohingya representatives could partic-ipate in the election. Thus, Rohingyas were cornered from all aspects and were deprived of their basic rights in their own land.

## 11. DISCUSSION AND CONCLUSION

The interviews reveal that the Rohingyas consider themselves as a different 'ethnic group' distinct both from other Muslim groups in Myanmar and from people in Bangladesh. To clarify where they came from, they often refer to their historical root, the living history of their ancestors. They consider Rakhine State as their ancestral homeland.

Historically, Myanmar has been a land where people from many places came and settled down. In fact, the ancient inhabitants of Burma are thought to be the descendants of Mongoloid tribes who mostly lived in upper Irrawaddy and interestingly there was no difference between Arakanese Maughs and Burmese except a little in language (Phayre 1883: 42). According to Harvey, historically Burma was mixture of tribes (Harvey 1925: 3). Harvey further pointed that the earliest inhabitants were Indonesians who were displaced by the Maughs and Tibet-Burmese tribes from western China. Subsequently the Maughs spread over the South to the near of Irrawaddy. Traditionally, the Tibet-Burmese tribes were known as Pyu, Kanran and Thet. The Thet are apparently recognized as Chins, and Kanran are now known as Arakanese (Harvey 1925: 3–4).

Amongst three Burmese attacks on Arakan Kingdom, the third attack in 1784–85 caused the greatest number of people to flow into Chattogram. During Bodawpaya occupation of Arakan in 1785, Burmese soldiers killed 20,000 Arakanese and several thousand fled to Chattogram as Walter Hamilton demonstrated that in 1802 he saw more than one hundred thousand Arakanese around twelve miles of Ramu in Cox Bazar (Akhanda 2013: 46). After the British annexation of Arakan in 1824 some of them went back to their homeland; however, the majority of them received British-Indian citizenship and settled permanently in Habrang, Chowdhurypara, Manikpur, Ramu, Cox Bazar town, Teknaf, Moheskhali, Hnila and some other places in Chattogram; moreover, right now they are living in those regions (Akhanda 2013: 55). While common argument is that some of Arakan people went from Chittagong, a deeper investigation reveals that the very composition of people in Chittagong was formed thorough the mix of people who migrated to Chittagong from Arakan during Burmese occupation of Arakan.

Moreover, Arakan Kingdom undertook two attacks on Chattogram and Bengal and one attack on Pegu while Chattogram was under the rule of Arakan. The Bengal rulers undertook two attacks on Ara-kan Kingdom. When British annexed Arakan, a massive labor flow ensued into Arakan for agricultural production. In addition, when British annexed Burma, the volume of migration increased manifolds. During the Second World War, the attacks and counterattacks between the British and Japanese forces, many people flew into Chattogram. All the events caused massive in- and outflow of people from one area to another. Moreover, there have been regular flows of people between the two adjacent areas.

During the Burmese invasion of Arakan, many Muslims prisoners were taken to upper Burma and they merged with Burmese community who are known as Zerbadees (*i.e*., Indian father and Burmese mother). Bamar Muslims were formed by conversion from Burmese Buddhists. Indian Muslim are descendants of Indian Muslims brought during the British annexation of Arakan and Burma (Yin 2005: 163). Pantay are Chinese Muslims who live along Burma-Chin border in the northern part of Burma and are considered as offspring of Nassaruddin who invaded Pagan. The Turks and Pathans are found in Mrauk-U. There are also Malay and Myay Du Muslims in Burma. Kaman or Kamanchi is the second largest Muslim group in Arakan, who are descendants of Mogul Emperor Shah Shuja and live in Ramree Island.

Due to the transformation of population composition over time, it is very difficult to separate Rohingyas from other Muslim ethnic groups in Arakan. In addition to the Maughs, there are four Muslim groups in Arakan, namely: Thambaikkya, Zerbadees, Kaman or Kamanchi and Rohingya (Karim 2000: 115) and there are eight Muslim groups in whole Myanmar, which include Bamar Muslims, Indian Muslims, Turks, and Pathans and Pantay (Mohajan 2018). The people living in Arakan/Rakhine have equal rights over the land regardless of their religion or ethnicity. The significance of the Rohingyas is no less important than any other ethnic group living in Myanmar. Thus, the Burmese authority's exclusionary policy towards the Rohingyas from other groups is not legitimate.

To summarize the findings, the territory of the Arakan/Rakhine state developed from the ashes of Arakan Kingdom over thousands of years, through a long history of independence, conquests, and counter conquests. Similarly, the communities and ethnic groups living in Arakan developed over thousands of years through numerous occasions of incoming and outgoing of the people. The Rohingyas have been living in Arakan for generations. Though the ethnic consciousness was created by the educated elite class in the 1950s, they referred to their ancestors' true historical legacy delineated in the article.

Even though soon after the establishment of Myanmar' independence, the government started accepting the Rohingyas in the state system reflecting territorial nationalism, unfortunately, with the advent of military government and subsequently adopted policies, the government started excluding the Rohingyas. The renaming of the Rakhine State and Burma are also part of exclusionary policies adopted by the military junta. The exclusion gradually denied their citizenship status, free movement, marriage and having children *etc*., which culminated in recent barbarous military eviction. Though from the perspective of Tatmadaw, its military operation of forcing out the Rohingyas from Rakhine state seems to be a success; it will surely be challenging to convert this short-term military success into a long-term political success (Minar 2019).

Thus, the development of ethnic nationalism among the Rohingyas may be considered within the push-pull framework, whereas the exclusionary practices of the Myanmar government pushed them on the one hand, the urge for survival pull them together to form a unity. However, the Rohingyas ambition in terms of ethnic nationalism is not aimed to any self-autonomy, let alone independent polity of their own, rather to get recognition and assurance of their basic rights, however, this was misinterpreted by the military government.

So, what future waits for the ousted Rohingyas? The Myanmar government initially agreed to accept bilateral initiatives with the government of Bangladesh to repatriate Rohingyas from Bangladesh which created optimism; however, there has not been any concrete progress so far. On the one hand, Myanmar government has been delaying implementing its agreed terms to repatriate Rohingyas; on the other hand, the UN's effectiveness to solve the Rohingya crisis seems inefficient since its initiatives have been blocked by China and Russia on multiple occasions (Minar 2018). The future of the Rohingyas seems uncertain now. Bangladesh remains optimistic, so is the world community; however, unfortunately there is no concrete framework for solution. Within such a scenario, the Myanmar government must perceive the long-term effect of denying Rohingyas basic rights and denying the history, which may be counterproductive to Myanmar governments' interests. If the government claims the territory of Rakhine State, then it should accept people living there. The sooner the government comes to realize this, the better it will be for Myanmar. Lastly, the United Nations Organization, the most successful international organization in the world history, should pursue to uphold its success in the case of the Rohingyas.

## NOTES

¹ Chin is a western state of Burma/Myanmar, mainly a mountainous region which is one of the least developed; Magway region is the administrative division of the central Burma with the second highest area among seven divisions of Burma, and Bago region is adjacent to Magway formerly known as Pegu and it has close historic relations with Arakan, adjacent to Bagothe Ayeyarwady which was previously called Irrawaddy, it also has contiguous relations with the proliferation of civilization in Arakan. Sittwe is a town in western Myanmar that was known as Akyab before the British annexation of whole Burma in 1885.

² Moulmein, now known as Mawlamyine, is the fourth largest city of Burma situated at 300 km south-east of Yangon.

³ A small group of thirty Burmese youth who went out of Burma during World War II and secretly obtain military training under Japanese Army to fight against the British Army on the battleground of World War II. They are usually called 'Thirty Comrades.'

⁴ Tatmadaw is the official name of Myanmar armed force including Army, Navy and Air force under the Ministry of Defence.